\documentclass[conference]{IEEEtran}
\IEEEoverridecommandlockouts
\usepackage{cite}
\usepackage{amsmath,amssymb,amsfonts}
\usepackage{array}
\usepackage{graphicx}
\usepackage{textcomp}
\usepackage{xcolor}
\usepackage{booktabs}
\usepackage{multirow}
\usepackage{makecell}
\usepackage{threeparttable}
\usepackage{algorithmicx}
\usepackage{amsmath}
\usepackage{amsthm}
\makeatletter
\long\def\@makecaption#1#2{%
  \vskip\abovecaptionskip
  \sbox\@tempboxa{{\normalsize #1: #2}}
  \ifdim \wd\@tempboxa >\hsize
    {\large #1: #2\par}
  \else
    \global \@minipagefalse
    \hb@xt@\hsize{\hfil\box\@tempboxa\hfil}
  \fi
  \vskip\belowcaptionskip}
\makeatother

\usepackage[ruled,vlined]{algorithm2e}

\def\BibTeX{{\rm B\kern-.05em{\sc i\kern-.025em b}\kern-.08em
    T\kern-.1667em\lower.7ex\hbox{E}\kern-.125emX}}
\def\etc{ \emph{et al.}}

\def\sv{EVeCA}

\begin{document}


\title{\sv: Efficient and Verifiable On-Chain Data Query Framework Using Challenge-Based Authentication}

\author{
            Meng Shen,~\IEEEmembership{Member,~IEEE},
            Yuzhi Liu, 
            Qinglin Zhao,~\IEEEmembership{Senior Member,~IEEE},       
            Wei Wang,~\IEEEmembership{Member,~IEEE},   
    \\
            Wei Ou,
            Wenbao Han,
		and Liehuang Zhu,~\IEEEmembership{Senior Member,~IEEE} 
		
		\IEEEcompsocitemizethanks{
		\IEEEcompsocthanksitem M. Shen, Y. Liu and L. Zhu are with the School of Cyberspace Science and Technology, Beijing Institute of Technology, Beijing 100081, China. (e-mail: \{shenmeng, liuyuzhi, liehuangz\}@bit.edu.cn).
            \IEEEcompsocthanksitem Q. Zhao is with the School of Computer Science and Engineering, Macau University of Science and Technology, Macau, China (e-mail: qlzhao@must.edu.mo).
            \IEEEcompsocthanksitem W. Wang is with the Beijing Key Laboratory of Security and Privacy in Intelligent Transportation, Beijing Jiaotong University, Beijing 100044, China (e-mail: wangwei1@bjtu.edu.cn).
            \IEEEcompsocthanksitem W. Ou and W. Han are with the School of Cyberspace Security (School of Cryptology), Hainan University, Haikou 570228, China (e-mail: \{ouwei, 994338\}@hainanu.edu.cn).
    }
}


\maketitle

\begin{abstract}


As blockchain applications become increasingly widespread, there is a rising demand for on-chain data queries. 
However, existing schemes for on-chain data queries face a challenge between verifiability and efficiency. Queries on blockchain databases can compromise the authenticity of the query results, while schemes that utilize on-chain Authenticated Data Structure (ADS) have lower efficiency.
To overcome this limitation, we propose an efficient and verifiable on-chain data query framework \emph{\sv}. In our approach, we free the full nodes from the task of ADS maintenance by delegating it to a limited number of nodes, and full nodes verify the correctness of ADS by using challenge-based authentication scheme instead of reconstructing them, which prevents the service providers from maintaining incorrect ADS with overwhelming probability. By carefully designing the ADS verification scheme, EVeCA achieves higher efficiency while remaining resilient against adaptive attacks. Our framework effectively eliminates the need for on-chain ADS maintenance, and allows full nodes to participate in ADS maintenance in a cost-effective way.
We demonstrate the effectiveness of the proposed scheme through security analysis and experimental evaluation. Compared to existing schemes, our approach improves ADS maintenance efficiency by about 20$\times$.

\end{abstract}


\section{Introduction}

\begin{figure}[t]
\centerline{\includegraphics[width =  .42\textwidth]{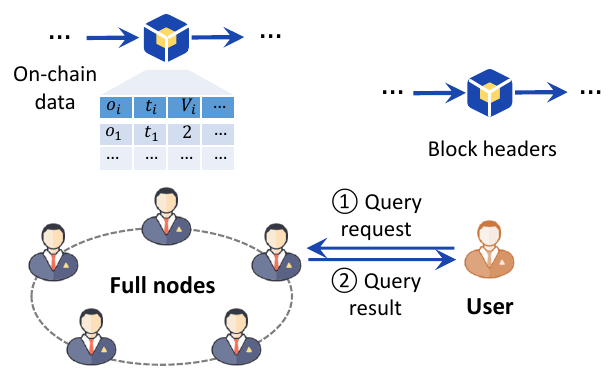}}
\caption{Data query process in a typical blockchain system. The user sends a query request to the full node, and the full node returns the query result to the user.}
\label{oncads}
\end{figure}

Blockchain, as a decentralized ledger technology, has offered a secure and transparent platform for interactions among mutually untrusted nodes.
Therefore, blockchain can be used in various applications\cite{10105989}, such as supply chain management\cite{tian2016agri}, IoT\cite{DBLP:journals/iotj/ShenTZDG19}, and healthcare\cite{DBLP:journals/network/ShenDZDG19}. In blockchain-based applications, users have a demand to query data from the blockchain for the purpose of transaction tracing, data auditing, and data analysis. For instance, in blockchain-based supply chain management systems, users (i.e., light nodes) may be required to query information about products with transaction time between a certain interval and an amount within a certain range, or products with descriptions that include certain keywords. As users only keep track of the block headers, they rely on full nodes to fulfill their querying needs, as shown in Fig.~\ref{oncads}.

The above form of on-chain data query has the following requirements.
(1) Verifiability: The correctness of query results must be verifiable, as full nodes are untrusted and may return incorrect query results\cite{DBLP:journals/internet/RenWW12} due to security vulnerabilities or the purpose of saving computational overhead. (2) Query efficiency: It is critical for users to obtain query results with low latency, while minimizing the computational cost of verifying the results. (3) Update efficiency: Due to the dynamic nature of on-chain data updates, full nodes need to minimize the computational cost associated with these updates.

Unfortunately, none of the existing works satisfies our aforementioned requirements.
(1) Research has been conducted on blockchain databases with high reading performance to achieve efficient on-chain data queries, such as EtherQL\cite{DBLP:conf/dasfaa/LiZYLZ17}, FlureeDB\cite{platz2017flureedb} and ECBC\cite{DBLP:conf/ictac/XuZKZZL17}. However, such data query schemes lack verifiability and users may be provided with incorrect query results.
(2) Another stream of approaches involves leveraging on-chain authenticated indexes, where full nodes maintain Merkle tree-based Authenticated Data Structures (ADS)\cite{DBLP:conf/sigmod/LiHKR06} in blocks to achieve verifiable query\cite{DBLP:conf/sigmod/0004ZX19, DBLP:conf/icdcs/DaiXYWCH020, DBLP:conf/icde/WangXZXPP22, DBLP:conf/icde/ZhuZJZY19, DBLP:journals/tsusc/YinZZW23}. When receiving a query request from users, the full node constructs proofs for query results based on ADS, thus providing users with verifiable query services. 
Due to the immutability of the blockchain, a one-size-fits-all ADS is needed to support dynamic query attributes\cite{DBLP:conf/sigmod/0004ZX19}. This requirement makes it challenging to simultaneously ensure both query efficiency and update efficiency. To support multi-dimensional queries, existing works construct attribute-aggregated indexes\cite{DBLP:conf/sigmod/0004ZX19, DBLP:conf/icpads/ZhuZJZ20}, or build a one-dimensional index on each attribute separately and take the intersection of the results of each attribute\cite{DBLP:conf/icde/WangXZXPP22}. Both of them have limited query performance. If a composite index is build for each combination of attributes, full nodes will incur significant update overhead.
A comparison of existing studies is shown in Table~\ref{tab:verifiable_query_methods}.

\begin{table*}[htbp]
\begin{threeparttable}
  \centering
\renewcommand{\arraystretch}{1.5}
  \caption{The Comparison with Existing Methods for Data Query in Blockchain Systems}
  \label{tab:verifiable_query_methods}
  \begin{tabular}{>{\centering\arraybackslash}p{2.25cm}|>{\centering\arraybackslash}p{2cm}|>{\centering\arraybackslash}p{2.2cm}|>{\centering\arraybackslash}p{1.6cm}|>{\centering\arraybackslash}p{1.6cm}|>{\centering\arraybackslash}p{2cm}|>{\centering\arraybackslash}p{1.5cm}|>{\centering\arraybackslash}p{1.5cm}}
    \toprule
    \multirow{2}{*}{\textbf{Techniques}} &\multirow{2}{*}{\textbf{\makecell{Representative \\ methods}}} & \multirow{2}{*}{\textbf{\makecell{ADS maintenance \\ efficiency\tnote{1}}}} & \multicolumn{3}{c|}{\textbf{Query efficiency\tnote{2}}}&\multicolumn{2}{c}{\textbf{Verifiability\tnote{3}}}\\
\cline{4-8}

 & & & \textbf{Range} & \textbf{Keyword} & \textbf{Multi-dimension} & \textbf{Soundness} & \textbf{Completeness} \\
    \hline
    \multirow{2}{*}{Off-chain database} & EtherQL\cite{DBLP:conf/dasfaa/LiZYLZ17} & \textbf{--} & $1\times \sim 2\times$ & $1\times \sim 2\times$ & $1\times \sim 2\times$ & $\times$ & $\times$ \\
    \cline{2-8}
    & VQL\cite{DBLP:journals/tpds/WuPGYX22} & \textbf{--} & $1\times \sim 2\times$ & $1\times \sim 2\times$ & $1\times \sim 2\times$ & \checkmark & $\times$ \\
    \hline
    \multirow{3}{*}{On-chain ADS\tnote{4}} & vChain\cite{DBLP:conf/sigmod/0004ZX19}  & $>10\times$ & $>10\times$ & $>10\times$ & $>10\times$ & \checkmark & \checkmark \\
    \cline{2-8}
    & LVQ\cite{DBLP:conf/icdcs/DaiXYWCH020} & $1\times \sim 5\times$ & \textbf{--} & $1\times \sim 2\times$ & \textbf{--} & \checkmark & \checkmark \\
    \cline{2-8}
    & vChain+\cite{DBLP:conf/icde/WangXZXPP22} & $>10\times$ & $1\times \sim 2\times$ & $1\times \sim 2\times$ & $>10\times$ & \checkmark & \checkmark \\
    \hline
    Off-chain ADS\tnote{5} & proposed \sv\ & $1\times \sim 5\times$ & $1\times \sim 2\times$ & $1\times \sim 2\times$ & $1\times \sim 2\times$ & \checkmark & \checkmark \\
    \bottomrule
  \end{tabular}
\begin{tablenotes}
\item [1] We use only the synchronization of on-chain data without building an index as a baseline. The conclusions are based on experimental results or theoretical analysis from the corresponding references.
\item [2] For the query type in each scheme, we use its corresponding verifiable query scheme for databases as the baseline scheme. Query efficiency represents the querying time required on the server side compared with the baseline.
\item [3] Verifiabiliy means that the user can verify the soundness (i.e., all answers satisfy the query expression and are not modified) and completeness (i.e., no valid data object is missing) of the results.
\item [4] On-chain ADS means that all full nodes maintain the ADS on the blockchain, which ensures the correctness of the ADS.
\item [5] Off-chain ADS means that a limited number of nodes maintain the ADS at local storage, where full nodes’ computational overhead is reduced, but requires additional methods to ensure ADS correctness.
\end{tablenotes}
\end{threeparttable}
\end{table*}


The key challenge is that ADS-based verifiable query schemes require, as a prerequisite, ensuring the correctness of the ADS itself. Typically, the full nodes are supposed to maintain the ADS to ensure its correctness through consensus (i.e., maintaining on-chain ADS), which incurs significant overhead for the full nodes. Therefore, the problem that needs to be addressed is: \emph{Can we manage to ensure the correctness of ADS in a cost-effective way?}

In this paper, we give an affirmative answer to the problem by proposing \emph{\sv}, a novel framework that supports efficient and verifiable on-chain data query. At its core, \sv\ frees full nodes from the task of ADS maintenance and query service provision by delegating it to a limited number of nodes, referred to as service providers (SPs). Correspondingly, the correctness of the ADS maintained by SPs is verified by the full nodes using challenge-based authentication scheme, which prevents the SPs from maintaining incorrect ADS with an overwhelming possibility. Based on the ADS verified by full nodes, users can further perform verifiable query processes with the SP.

In \sv, the SPs extract on-chain data, construct ADS and provide query services to users. Before user queries, the full nodes first verify the correctness of the ADS maintained by SPs. Specifically, the challenge node (one of the full nodes) constructs a detecting token (i.e., query expressions with known results) as a challenge and sends it to the SP. Then, the SP publishes the query result and proof of the detecting token, and the full nodes verify the correctness of the published result and the validity of the proof. By carefully designing the ADS verification scheme, our approach can resist adaptive attacks, and any attempts by maintaining incorrect ADS from SPs will be detected with a high probability. Also, by enabling existing Merkle tree-based ADS to be applied to the on-chain data query scenario, our framework can support multiple query types. \sv\ effectively eliminates the need for full nodes to maintain on-chain ADS, and allows them to participate in ADS maintenance at minimal cost.

In summary, we summarize our contributions as follows:

\begin{itemize}
\item We propose \sv, a novel framework for efficient and verifiable on-chain data query. In \sv, the task of ADS maintenance is delegated from full nodes to a limited number of SPs, and the correctness of ADS is verified by full nodes without reconstruction. Our method effectively eliminates the need for on-chain ADS maintenance.
\item We design a challenge-based authentication scheme for full nodes to verify the correctness of the ADS maintained by SPs. This scheme achieves ADS verification with minimal cost while remaining resilient against adaptive attacks conducted by SPs.
\item We conduct a security analysis to prove the security of \sv. Also, we carry out comprehensive performance evaluations on detecting rate, ADS maintenance cost and query cost. Experimental results show that our scheme outperforms the existing schemes by about 20$\times$.
\end{itemize}


\textbf{Organization. }The rest of the paper is organized as follows. We describe the background of verifiable query for on-chain data in Section II. Then, we define the verifiable query problem in Section III, and describe the detailed design of the proposed scheme in Section IV. We discuss the scenario of adaptive attacks in Section V. Next, we exhibit the security analysis in Section VI, evaluate the proposed method through experiments in Section VII. Finally, We summarize the existing works in Section VIII and conclude this paper in Section IX.

\section{Background}

In this section, we present the background and problem definition of verifiable queries for blockchain.


\subsection{Authenticated Data Structure}



Authenticated Data Structure (ADS)\cite{DBLP:conf/sigmod/LiHKR06} is a type of data structure that allows users to verify the correctness and integrity of the stored data. These structures are designed to provide cryptographic proofs that can be used to confirm that the data has not been tampered with and that it is consistent with the expected values. ADS is commonly used in distributed systems to ensure that data remains secure and free from tampering.

One of the most widely used types of ADS is the Merkle tree-based ADS. A Merkle tree\cite{DBLP:conf/crypto/Merkle89} is a binary tree in which each non-leaf node is labeled with the hash of the labels or values of its child nodes. The leaves of the tree contain the actual data, and the root of the tree is a single hash value that represents the entire dataset. This root hash is often referred to as the Merkle root.

\begin{figure}[htbp]
\centerline{\includegraphics[width =  .4\textwidth]{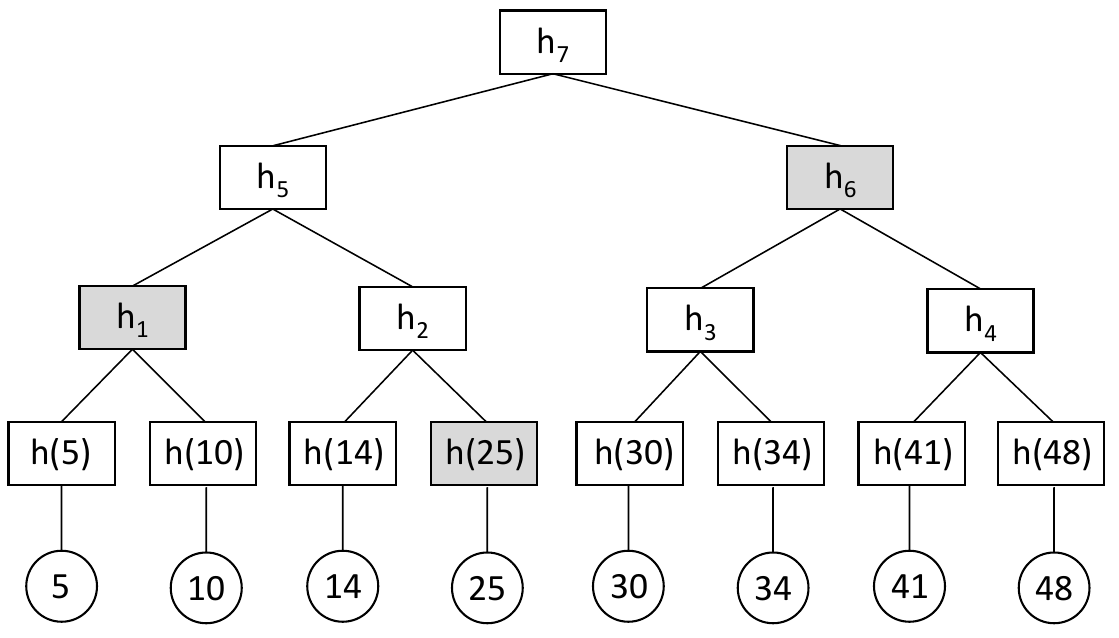}}
\caption{Example of a Merkle tree}
\label{mt}
\end{figure}

\emph{\textbf{Example 1. (Merkle tree)} In a Merkle tree shown by Fig.~\ref{mt}, the proof for the existence of element 14 is as follows: 
(1) The full node sends the verification data structure $(h(25),h_1,h_6)$ to the user; 
(2) The user calculates $h_2,h_5,h_7$ in turn according to $(h(25),h_1,h_6)$; 
(3) The user compares $h_7$ and the locally stored root hash value, and if it is equal, the verification is successful.}

To achieve query efficiency, Merkle trees can be integrated into data structures that are optimized for querying. For example, a B-tree or a balanced search tree can be augmented with Merkle tree properties, where each node also stores a hash of its children, forming a Merkle B-tree\cite{DBLP:conf/sigmod/LiHKR06} structure. This integration allows for efficient query while still providing the cryptographic guarantees of a Merkle tree. In this paper, the ADS all refer to Merkle tree-based index structures that support efficient querying.

\begin{figure*}[htbp]
\centerline{\includegraphics[width =  1.0\textwidth]{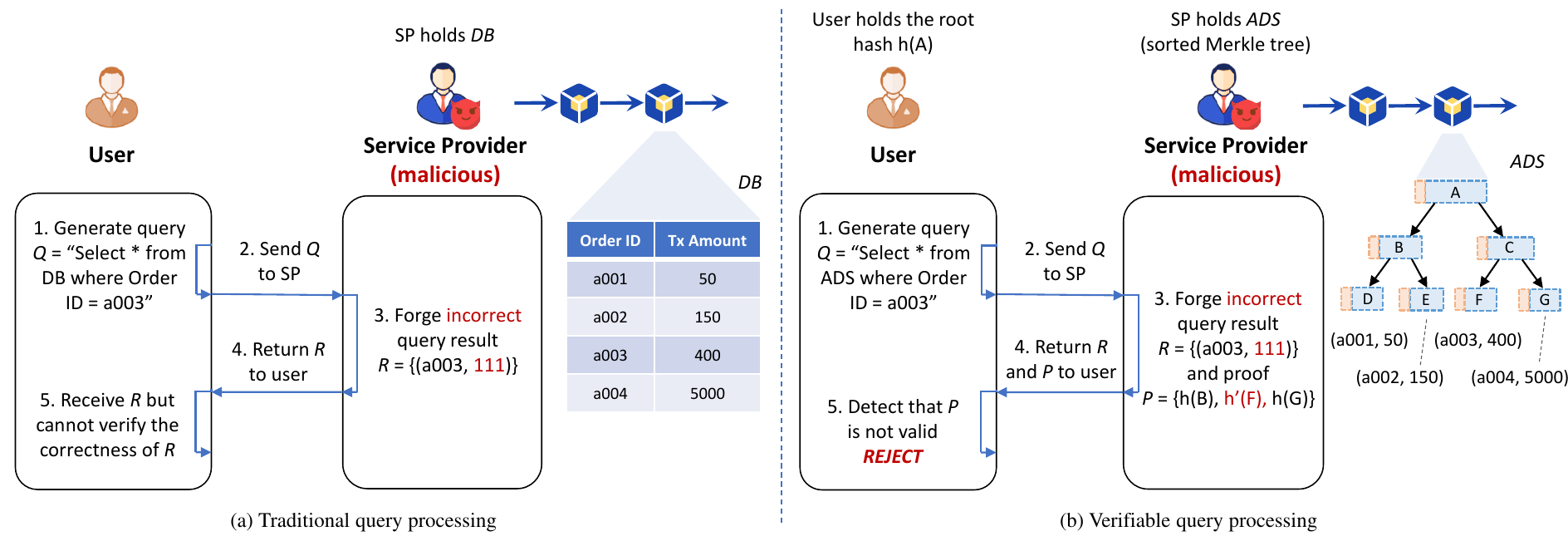}}
\caption{An example illustrating verifiable query. In the traditional query processing (a), the user cannot verify the correctness of the query result. But in the verifiable query processing (b), the SP must return the query proof, with which the user can verify the correctness of the query result.}
\label{vq}
\end{figure*}

\subsection{Verifiable Query for On-chain Data}

Verifiable query aims to provide a method for users to verify the query results obtained from untrusted SPs. 
In a typical on-chain data verifiable query scheme, the full nodes maintain the on-chain data in the ADS, while the light nodes are query users and only keep track of the block headers. When receiving a query request sent by light nodes, the full node returns the query result as well as the proof. Given a set of on-chain data $D$ containing $|D|$ elements, the definition of verifiable query includes the following elements:

$\delta \leftarrow Gen\_Digest(D)$: This function takes $D$ as input and output the digest $\delta$ of dataset $D$.

$(R, P)\leftarrow Query(D, Q)$: This function takes as input $D$ and query $Q$, and output query result $R$ and proof $P$.

$\{0, 1\}\leftarrow Verify(R, P, \delta)$: This function takes as input $R$, $P$ and the digest $\delta$, and output 1 if $R$ is verified correct, otherwise output 0.

\emph{\textbf{Example 2. (Verifiable query)} Assuming that the SP maintains a dataset with four data objects, in the absence of a verifiable query mechanism, which is shown in Fig.~\ref{vq}(a), the user queries the data with $Order ID=a003$. If the SP returns a forged result (a003, 111), the user cannot detect it. In the case of introducing the verifiable query mechanism, which is shown in Fig.~\ref{vq}(b), the SP organizes the data in a sorted Merkle tree. If the SP attacks, the user can calculate the hash step by step according to the proof returned by the SP, i.e., $H(h_B||H(h'_F||h_G))\neq h_A$, and detect that the calculated root hash is incorrect.}

Common types of verifiable on-chain data queries include the following. We next briefly introduce their definitions and common methods.

\emph{Range query. }Range query is a type of query operation that is based on a certain attribute or value range. It is in the form of $Q=<[q^l, q^u]>$, where $q^l$ is the lower bound  and $q^u$ is the upper bound. Range queries include query of the timestamp of data objects within a certain range, or a numerical attribute of data objects that satisfy certain conditions. For verifiable range queries, a common approach is to construct a Merkle B-tree\cite{DBLP:conf/sigmod/LiHKR06} on the attribute to be queried, which is a kind of efficient ADS for range queries.

\emph{Object query. }Object query is a special case of range query, which is in the form of $Q=<[q^l, q^u]>$ with $q^l=q^u$. The methods used for object query can be the same as those used for range query.

\emph{Multi-dimensional query. }Multi-dimensional query is the process of querying data objects that satisfy certain conditions across multiple attributes of the data. It is in the form of $Q=<[q[1]^l, q[1]^u], [q[2]^l, q[2]^u], ..., [q[d]^l, q[d]^u]>$, where $d$ is the number of query attributes. For multi-dimensional queries, common methods include building composite indexes using MB-trees on combinations of attributes, or utilizing high-dimensional ADS such as Merkle R-trees\cite{DBLP:conf/icde/YangPPK08}.

\emph{Keyword query. }Keyword query is a search operation where users input specific words or phrases to find relevant information. It is in the form of $Q=<w_1, w_2, ..., w_n>$, where $n$ is the number of keywords contained in the query. To process verifiable keyword queries, a commonly used data structure is the Merkle inverted index\cite{DBLP:conf/icde/ZhangXWXC21}.




\section{Problem Formulation} \label{sec:prob}

This section presents the system model of the proposed \sv, as well as the threat model and design goals.

\subsection{System Model}

\begin{figure}[t]
\centerline{\includegraphics[width =  .51\textwidth]{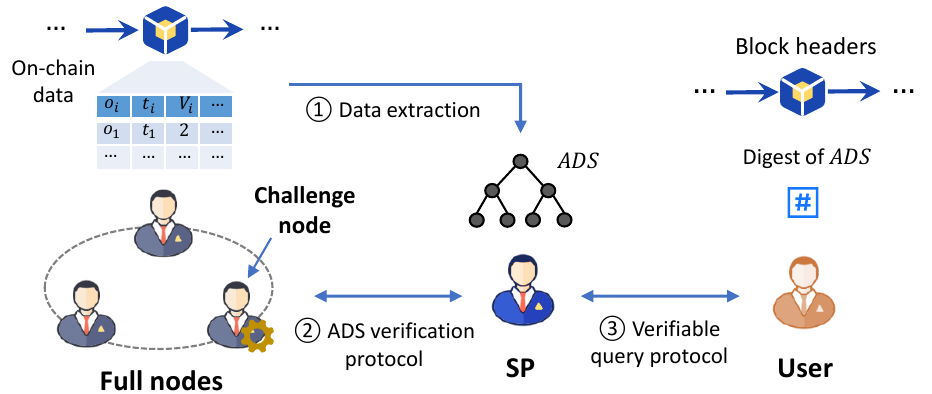}}
\caption{System model of \sv. The task of constructing the ADS is assigned to a limited number of SP nodes, while full nodes are responsible for verifying the correctness of the ADS. Finally, users initiate verifiable query requests to the SPs that have passed the verification.}
\label{sysmodel}
\end{figure}

To enable efficient and verifiable queries for on-chain data, we build a new model that contains the following entities shown in Fig.~\ref{sysmodel}: service provider (SP), query user, full node and challenge node.


\emph{Service provider (SP).} The SPs are nodes with high storage and computing resources. They maintain ADS locally, extracting on-chain data and update their local ADS after a certain number of blocks is created. They provide query services to the users.

\emph{Query user.} The query users are light nodes that keep track of the block headers. They send query requests to the SP, and verify the query result according to the proof from the SP.

\emph{Full node.} The full nodes maintain a complete copy of the blockchain and validate new transactions and blocks. They are responsible for verifying the correctness of the ADS maintained by the SP. 

\emph{Challenge node.} Challenge nodes are a specialized subset of full nodes, possess the capability to directly send detecting tokens to the SP as challenges.

\begin{figure*}[t]
\centerline{\includegraphics[width = 1.0\textwidth]{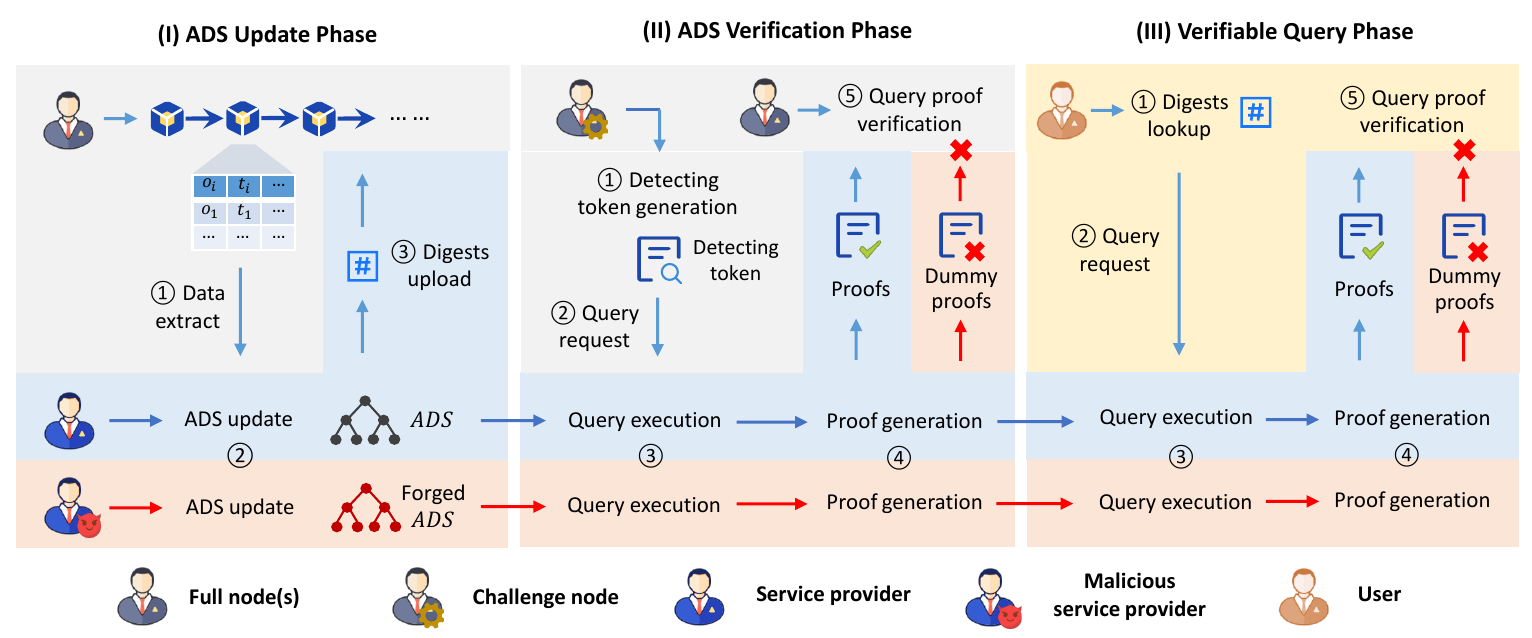}}
\caption{System overview of \sv. In the ADS update phase, SPs extract on-chain data, update their local ADS and upload the digest of ADS to the blockchain. In the ADS verification phase, the challenge node generates a detecting token and request to SP, and full nodes verify the correctness of the returned result and proof. In the verifiable query phase, users send query request to the SP and verify the returned results based on the digest of ADS.}
\label{framework}
\end{figure*}

The data queried by users can be modeled as a sequence of data objects. Each of the objects can be represented as $\{ts, a_1, a_2, ..., a_n\}$, where $ts$ is the timestamp of the data object, $a_i$ is the $i$-th attribute of the object. 

\subsection{Threat Model}

The SPs are assumed to be untrusted parties. SPs may have program failures, security vulnerabilities and other security issues. Meanwhile, SPs have the need to save computing and storage resources\cite{DBLP:journals/internet/RenWW12}. Therefore, the SP may maintain incorrect ADS.
We classify security threats caused by SPs into the following two points.
(1) Data omission: The SP omits data from the maintained ADS.
(2) Data misplacement: The SP misplaces data in ADS.

It is worth noting that existing works\cite{DBLP:conf/sigmod/0004ZX19, DBLP:conf/icdcs/DaiXYWCH020, DBLP:conf/icde/WangXZXPP22, DBLP:conf/icde/ZhuZJZY19, DBLP:journals/tsusc/YinZZW23} have studied ADS supporting verifiable queries, which can verify the correctness of the query results under the premise of ensuring the correctness of ADS. Therefore, this aspect is beyond the scope of consideration in our work.

We hypothesize the potential collusion between challenge nodes and SPs, but ensuring the randomness of detection tokens can counter such collusion effectively (refer to \cite{kiayias2017ouroboros, clink} for specific methods). Therefore, this scenario is also beyond the scope of this work.



\subsection{Design Goals}


To meet the demand for efficient and verifiable on-chain data query, \sv\ has the following design goals:

\textbf{Verifiability. }To ensure users to receive accurate query results, the verifiability of the query scheme must be guaranteed. Verifiability means that the user can verify the soundness (i.e., all answers satisfy the query expression) and completeness (i.e., no valid data object is missing) of the results.

\textbf{Query efficiency. }Considering the increasing volume of data and the need for real-time data analysis, query efficiency must be guaranteed. Query efficiency represents the time overhead for query on the server side and verification on the user side.

\textbf{ADS update efficiency. }Due to the high-frequency update nature of on-chain data, the efficiency of ADS updates is crucial. ADS update efficiency represents the time overhead required for ADS updates on full nodes.


\begin{table}
\centering
\renewcommand{\arraystretch}{1.3}
\caption{Notations Used in This Paper}
\vspace{4pt}
\label{tab:notation}
\begin{tabular}{|c|c|}
\hline
\textbf{Notations} & \textbf{Description} \\
\hline
$k$ & Fraction of omitted data objects \\
\hline
$\lambda$ & Security parameter \\
\hline
$D$ & Newly updated on-chain data set \\
\hline
$|D|$ & Data update size \\
\hline
$\mathcal{A}_D$ & ADS generated from the dataset $D$ \\
\hline
$\delta$ & The digest of ADS \\
\hline
$h_i$ & The hash field of treenode $i$ \\
\hline
$Q_u$ & Query expression requested by the user \\
\hline
$Q_d$ & Detecting token generated by challenge node \\
\hline
$R_u$ & Query result of $Q_u$ returned by the SP \\
\hline
$R_d$ & Query result of $Q_d$ returned by the SP \\
\hline
$V_R$ & Verification object of query result $R$ \\
\hline
$S$ & Omitted data set by SP \\
\hline
$P_{suc}$ & Attack success rate of SP \\
\hline
$P_{d}$ & Detecting success rate of full nodes \\
\hline
\end{tabular}
\end{table}

\section{The Proposed \sv} \label{sec:sch}

In this section, we introduce the proposed scheme for efficient and verifiable on-chain data query.

\subsection{Overview}


\textbf{Basic idea. }
The basic idea of this paper is to verify the correctness of the ADS by full nodes using a challenge-based verification scheme instead of reconstruction. This approach effectively eliminates the need for full nodes to maintain on-chain ADS. The insight is that when the SP omit/misplace some data objects in the ADS, he cannot construct valid proof for queries containing the omitted/misplaced data objects. Also, by conducting sampling inspections on only a small portion of the data, it can detect a certain amount of data omissions/misplacement with an overwhelming probability. Based on the above insights, our approach is to send a detecting token (i.e., query expression with known results) to the SP, and determine the correctness of the ADS maintained by the SP by observing the results returned by the SP on the detecting token.

The system process of \sv\ is divided into \emph{ADS update phase}, \emph{ADS verification phase}, and \emph{verifiable query phase}, which is shown in Fig.~\ref{framework}.

\textbf{ADS update phase. }When on-chain data is updated, SPs first synchronize on-chain data to local storage. Then, they update their maintained ADS.
At the end of each round of data updates, SPs put the digest (root hash) of their ADS on the blockchain as prior knowledge for verifiable queries.
The specific construction method of ADS can refer to the method adopted by the related work on verifiable queries in outsourced database scenarios.


\textbf{ADS verification phase. }In the ADS verification phase, the challenge node constructs detecting token (i.e., query expression with known result) and sends it to the SP.
Then, the SP returns the query result and its proof.
Lastly, all the full nodes verify: (1) the correctness of the query result on the detecing token, and (2) the validity of the proof. If at least one of the above verification fails, it is determined that the ADS maintained by the SP is incorrect.

\textbf{Verifiable query phase. }In the verifiable query phase, users can send the query requests to the SP, and verify the query result proof returned by the SP. Since the correctness of the ADS maintained by the SP has been verified by full nodes, the security of user queries can be guaranteed with an overwhelming probability.

\subsection{ADS update phase}

In the ADS update phase, the SP extracts on-chain data, and updates their maintained ADS for a certain time interval.
Afterwards, SPs upload the digest (i.e., the root hash) of their ADS onto the blockchain. This serves as prior knowledge for verifiable queries, which allows for the verification of query results from the SP.

We optimize the hash calculation method of ADS. When calculating the hash field of each node, we concatenate the hash of its child nodes and the hash summation of all data objects contained in the child nodes. Suppose that the child hashes of node $n$ are $h_1, h_2, ..., h_i$, and data objects below the node are $o_1, o_2, ...o_j$, the hash of the node $n$ can be calculated as $h_n=H(H(o_1)+H(o_2)+...+H(o_j)||h_1||h_2||...||h_i)$.
Aside from the root hash of ADS, the SP is also supposed to publish $\sum_{i=1}^{|D|} H(o_i)$. As full nodes also maintain all the on-chain data, they can verify the correctness of the hash summation $\sum_{i=1}^{|D|} H(o_i)$, thus determine whether the data sets they contain are complete.

\begin{figure}[htbp]
\centerline{\includegraphics[width =  .46\textwidth]{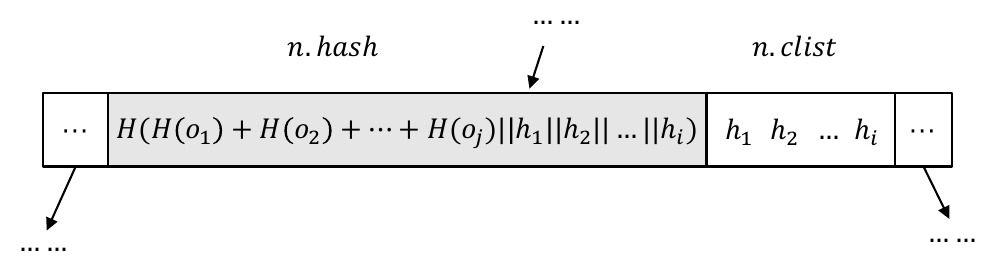}}
\caption{The node structure of the ADS. When calculating the hash of each node, we concatenate the hashes of its child nodes with the sum of the hashes of all data objects contained in those child nodes.}
\label{ads}
\end{figure}

When the SP extracts the data from the blockchain, he inserts the data in recently updated blocks into the currently established ADS. To reduce the overhead of maintaining ADS for the SP, we set the maximum size of ADS. When the ADS exceeds this size, the SP places the data in a newly generated ADS. The process of the ADS update phase is illustrated in Algorithm \ref{algo:updatephase}.

It should be noted that the uploaded digests from the SP are \emph{not necessarily reliable}, because the SP may omit/misplace data in the ADS and upload its root hash. Therefore, to ensure the correctness of the query result, further verification approaches are needed.


\begin{algorithm}[t]
\renewcommand{\baselinestretch}{1.15}
\small
  \label{algo:updatephase}
\LinesNumbered
  \SetAlgoLined
  \KwIn{The blockchain $BC$, newly generated on-chain data $D$, the authenticated indexes $(\mathcal{A})$ SP maintain locally, max size of ADS $maxsize$}
  \KwOut{The authenticated indexes $(\mathcal{A})^{new}$ on SP side after data update}

\textbf{\emph{/* At SP */}}

  \For{a certain time interval}{

    	Synchronize $D$ from the blockchain

  \eIf{the size of current ADS reaches $maxsize$}{

    	Generate a new ADS $\mathcal{A}^{new}_{D}$

    	Calculate the digest $\delta^{new}_D=Gen\_Digest(\mathcal{A}^{new}_{D})$

    	Put $\delta^{new}_D$ on $BC$}{
	
    	Update local database and ADS $\mathcal{A}_{D}$

    	Calculate the digest $\delta_D=Gen\_Digest(\mathcal{A}_{D})$

    	Update $\delta_D$ on $BC$}
  }
  
  \caption{ADS update phase}

\end{algorithm}


\subsection{ADS Verification Phase}


In the ADS verification phase, the full nodes verify the correctness of the ADS maintained by the SP. We design a challenge-based verification method. The challenge node sends a detecting token to the SP, and the full nodes verify whether the SP returns correct query results on the detecting token.


We assume that dishonest SPs save overhead by a certain percentage. We can approximate that when a SP tries to perform data omission attack, the fraction of the cost they are trying to save is the fraction of the data they omit. Intuitively, the more data is included in the detecting token, the higher the probability that a malicious SP will be detected. The challenge node first calculates the number of data objects that should be included in the query expression, and then constructs the qualified query expression. The process of the ADS verification phase is illustrated in Algorithm \ref{algo:veriphase}.

The statistical security parameter is denoted by $\lambda$. The security requirement aims to ensure that the SP's probability of successfully cheating is limited to at most $2^{-\lambda}$.
The challenge node specifies the parameters $k$ (the fraction of the SP omits) and $\lambda$, and calculates the value of $N$. Specifically, for every choice of $k$, the challenge node searches until finding the smallest $N$ that satisfies $P_{suc} \leq 2^{-\lambda}$.

\textbf{Detecting token size calculator. }Suppose the number of data objects included in the query expression with known results is $N$, the total amount of data is $X$, and the number of data objects missed by the SP is $x$.


We assume that the SP omits/misplaces a fraction $k$ of the maintained data, the probability that the attack succeeds is

\begin{equation}
P_{suc}=(1-\frac{kN}{x})^x
\end{equation}

Since we assume a large amount of data on the chain, the problem can be viewed as the problem of finding the limit as $x$ tends to infinity, i.e.

\begin{equation}
P_{suc}=\lim_{x\to\infty}(1-\frac{kN}{x})^x=\frac{1}{e^{kN}}
\end{equation}

It follows that the number of data objects contained in the inserted query expression is

\begin{equation}
N=\frac{1}{k}\ln_{}{\frac{1}{1-P_{d}}}
\end{equation}

\textbf{Theorem 1. }When the SP randomly omits/misplace the fraction $k$ of the data, to ensure a detection success rate of $P_d$, at least $N=\frac{1}{k}\ln_{}{\frac{1}{1-P_{d}}}$ data objects should be included in the detecting token.

To illustrate Theorem 1, we consider a specific scenario. We assume that each verification token for detection contains 1\% of the currently updated data and conduct multiple detections. Suppose we set the desired detection success rate $P_d$ to 99\% and the fraction of omitted/misplaced data $k$ to 0.1\%. Using the formula from Theorem 1, we can calculate that approximately 460 data objects need to be included in the detecting token. If the total data volume in an update is 200,000, then 460 data objects represent about 2.3\% of the total data, which means that 3 detections (each containing 1\% of the data) are required to achieve the desired detection success rate of 99\%.

\textbf{Random ranges selection. }After determining the volume of the detecting token, the challenge node randomly selects several ranges of the same volume so that their total volume is $N$. The volume of each range should be no less than the fan-out of the ADS. This design acts as an optimization for adaptive attacks which is further discussed in Section \ref{sec:adap}. Finally, the challenge node sends the detecting token to the SP as a query request.

\textbf{Proof verification. }To verify the correctness of the ADS maintained by the SP, the full nodes verifies the correctness of the query result of the detecting token and the validity of the proof. The full nodes first compare the query results published by the SP with the on-chain data they maintain. This comparison allows them to determine whether the query result on the detection token provided by the SP is correct. Additionally, full nodes verify that the proof of the query results published by the SP is consistent with the provided query result. If the above two parts of verification are passed, the verification is accepted, otherwise, the verification is rejected.

\begin{figure}[t]
\centerline{\includegraphics[width =  .46\textwidth]{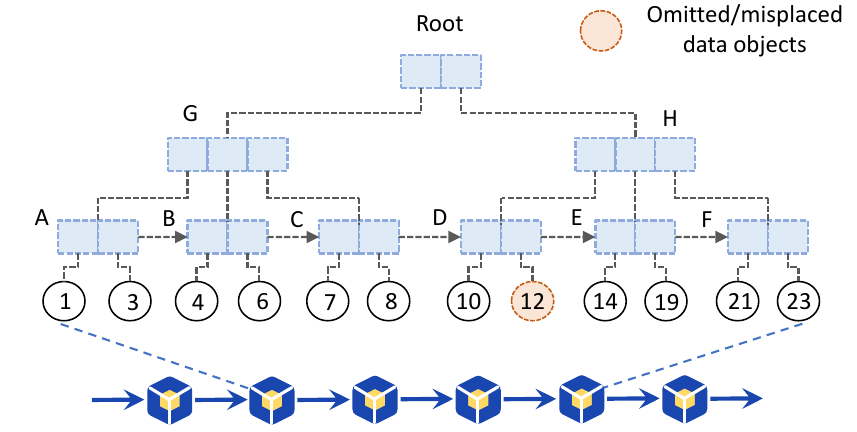}}
\caption{Example of verifiable query scheme. Assume that the SP maintains an MB-tree for on-chain data within a certain time range and omits/misplaces the data object \{12\}. If the detection expression includes the data object \{12\}, the SP cannot construct a valid proof. See Example 3 for details.}
\label{examplequery}
\end{figure}

\emph{\textbf{Example 3. (Detection of data omission/misplacement)} In the example of Fig.~\ref{examplequery}, supposed the detection expression of the challenge node is $Q'=[10, 14]$, the SP first finds out the query result $R=\{10, 12, 14\}$, and constructs the proof of the result $V_R=\{h(19), h(F), h(G)\}$. Then, the full nodes verify:}

\emph{(1) whether the query result $R$ is correct;}

\emph{(2) whether $V_R$ is valid by calculating $h(D)$, $h(E)$, $h(H)$ and check whether $h(Root)$ is correct.}

\emph{If the above two verifications are passed, the verification is accepted, otherwise the verification is rejected.}

\emph{Supposed the SP omits/misplaces data object $\{12\}$, then if the SP attempts to deceive the detection, he will return the correct result $\{10, 11, 12\}$ on the detection expression. But in this case, the SP is not able to construct the query proof consistent with the root hash he uploaded previously.}

\begin{figure}[t]
\centerline{\includegraphics[width =  .49\textwidth]{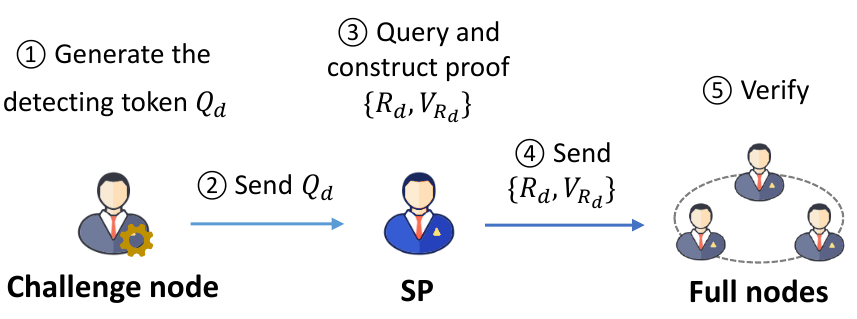}}
\caption{System process of ADS verification phase. The challenge node sends a query request to the SP, then the SP discloses the query result and the proof. Finally, the full nodes verify the correctness of the result and the validity of the proof.}
\label{querystage}
\end{figure}



\subsection{Verifiable Query Phase}


In the verifiable query phase, the user first generates query requests and send them to the SP. Then, the SP generates the query proof based on the digest. The specific method of generating proofs can be referred to the related work on verifiable queries in outsourced database scenarios, such as MB-tree for range queries and Merkle inverted indexes for keyword queries. When verifying the query results, the user verifies whether the proof provided by the SP aligns with the previously uploaded digest. The process of the verifiable query phase is illustrated in Algorithm \ref{algo:queryphase}.

It is worth noting that the digest of ADS uploaded by the SP does not serve as reliable prior knowledge for the users. However, due to the detection process of ADS verification phase, the behavior of SPs maintaining incomplete data is highly likely to be detected.


\emph{\textbf{Example 4. (Detection of data omission in query result)} In the example of Fig.~\ref{examplequery}, consider a range query $Q=[5, 8]$, the SP first find out the query result $\{6, 7, 8\}$. and construct the proof of the result $V_R=\{h(4), h(A), h(H)\}$. Then, the query user verify whether $V_R$ is valid by calculating $h(B)$, $h(C)$, $h(G)$ and check whether $h(Root)$ is correct.}

\emph{Supposed that the SP omits data object \{6\}, the user can detect that \{4\} \{7\} is not an adjacent leaf node according to the properties of the MB tree.}

\begin{algorithm}[t]
\renewcommand{\baselinestretch}{1.15}
\small
\label{algo:veriphase}
\LinesNumbered
	\SetAlgoLined
	\KwIn{The blockchain $BC$, security parameter $k$ and $\lambda$, newly updated data set $D$, newly updated digest of ADS $\delta$}
	\KwOut{ $Accept$ if updated ADS is verified correct; otherwise, $Reject$}
	
	\textbf{\emph{/* At challenge node */}}





	Calculate the number of data objects $N$ included in the query expression that satisfies $P_{suc} \leq 2^{-\lambda}$

	Construct query expression with known result $Q_d$ based on $N$ and send $Q_d$ to SP

	\textbf{\emph{/* At SP */}}

	Get $Q_d$ from challenge node

	Calculate $(R_d, V_{R_d})=Query(D, Q_d)$ and put $(R_d, V_{R_d})$ on $BC$

	\textbf{\emph{/* At full nodes */}}
 
        Get returned results $R_d$ and $V_{R_d}$ from $BC$
	
	Get the related data $R_d$ from on-chain data that satisfy query expression $Q_d$

   

	
	\eIf{$R_d$== $R_0$}{
		
		\eIf{$Verify(R_d, V_{R_d}, \delta)==1$}{
			return $Accept$
		}{			
			return $Reject$
		}
	}{
		
		
		return $Reject$
	}	
	\caption{ADS Verification Phase}
\end{algorithm}

\begin{algorithm}[t]
\renewcommand{\baselinestretch}{1.15}
\small
\label{algo:queryphase}
\LinesNumbered
	\SetAlgoLined
	\KwIn{The blockchain $BC$, on-chain data set $D$, digest of ADS $\delta$}
	\KwOut{ $Accept$ if query result is verified correct; otherwise, $Reject$}
	
	\textbf{\emph{/* At user */}}

	Send query request $Q_u$ to SP

	\textbf{\emph{/* At SP */}}

	Get $Q_u$ from user

	Calculate $(R_u, V_{R_u})=Query(D, Q_u)$ and return $(R_u, V_{R_u})$ to the user

	\textbf{\emph{/* At user */}}

	Get returned results $R_u$ and $V_{R_u}$ from the SP
		
	\eIf{$Verify(R_u, V_{R_u}, \delta)==1$}{
		return $Accept$
	}{			
		return $Reject$
	}	
	\caption{Verifiable Query Phase}
\end{algorithm}

\section{Adaptive Attacks} \label{sec:adap}

In this section, we introduce the adaptive attacks that can be performed by SPs and discuss the robustness of \sv\ to adaptive attacks.

\emph{\textbf{Attack 1.} The SP performs data omission/misplacement attack within concentrated ranges of a specific attribute.}

\begin{figure}[t]
\centerline{\includegraphics[width =  .5\textwidth]{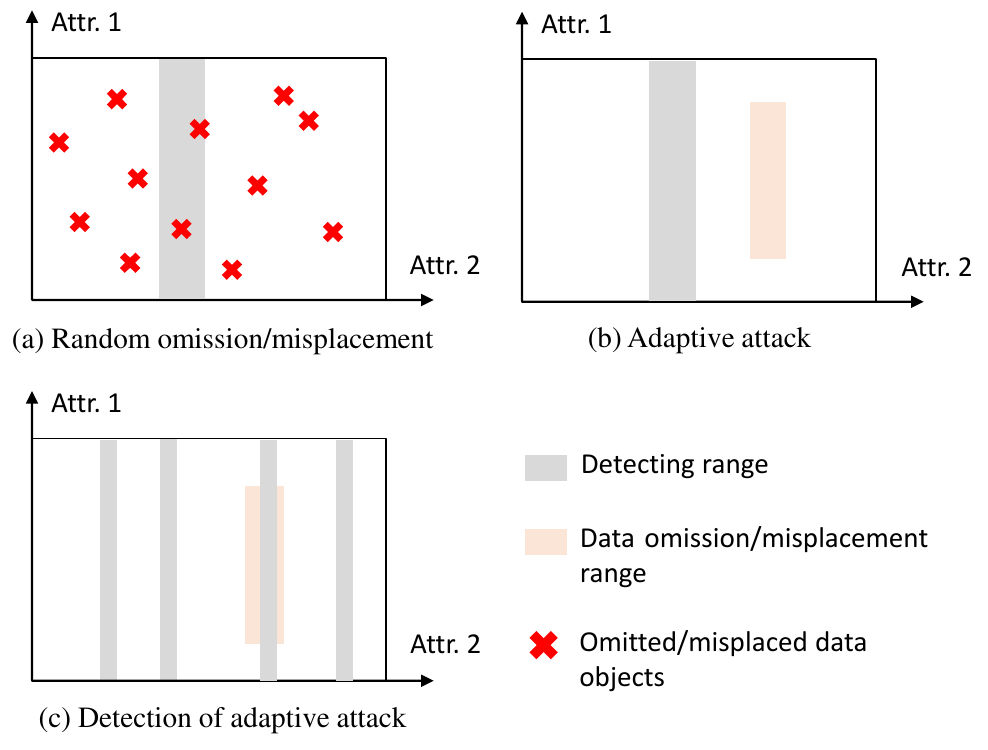}}
\caption{Adaptive attack of the SP. Random data omission/misplacement can be detected with an overwhelming
possibility (a), while data omission/misplacement in concentrated ranges can reduce the detection rate (b). Splitting the detecting token into more ranges can still maintain the detection rate at a high level (c).}
\label{adaptive}
\end{figure}

When the SP randomly omits or misplace data, as shown in Fig.~\ref{adaptive}(a), it will be detected with an overwhelming possibility. However, this assumption is not necessarily true, because SPs may conduct targeted evasion attacks after knowing the detection scheme. The SP can select a number of ranges and perform data omission attacks in the selected ranges. When the detection ranges and the data omission ranges do not intersect, it means that the detection has failed, as shown in Fig.~\ref{adaptive}(b). As an illustration, we represent the data space with two attributes as a two-dimensional plane, where the horizontal and vertical coordinates represent one attribute (e.g. timestamp and transaction amount).




To improve the detection rate of attacks, our approach is to split the detecting token into more ranges (introduced in Section \ref{sec:sch}) to enhance the detection capability, as shown in Fig.~\ref{adaptive}(c). This practice will increase the verification overhead of the queries to some extent, but even so, our solution still has significant advantages in ADS verification efficiency.

The adaptive attacks can result in a decrease in the detection rate. Nevertheless, we posit that attackers engage in adaptive attacks to reduce computational expenses, yet they only omit/misplace an exceedingly small quantity of data, which holds no significance for them. The omission/misplacement of such amount of data is more prone to occur in unforeseen circumstances (program malfunctions, etc.)

Due to the design of hash accumulation (introduced in Section \ref{sec:sch}), data omissions by the SP can be directly detected. Although SPs may attempt to evade our solution with evasion attacks, our approach still effectively counters such attacks. We illustrate this with the following attack as an example.

\emph{\textbf{Attack 2.} The SP omits an amount of data objects and forges the cumulative sum of hash of the node during index construction so as to deceive the challenge node. We use the following example for illustration.}

\emph{\textbf{Example 5. (Explanation of Attack 2)} In the example of Fig.~\ref{examplequery}, supposed the SP omits data object \{12\}, then if he attempts to cheat the full nodes, he may calculate the hash of node G as $h_G=h(H(10)+H(12)+H(14)+H(19)+H(21)+H(23)||h(D)||h(E)||h(F))$, though data object \{12\} is in fact not contained in the maintained ADS. In this case, if the omitted data objects are not queried by challenge node, the attack cannot be detected.}

We have the insight that when such misindexing exists in the ADS, there always exists a query for which the SP cannot construct a valid proof. Therefore, when the number of users querying is sufficiently large, the detection success rate is close to 100\%. Therefore, \emph{Attack 2} is invalid.

Moreover, we consider that data omission due to unexpected conditions tend to be randomly distributed in attributes other than timestamps under the Missing-at-random (MAR) assumption\cite{DBLP:conf/ecml/Jaeger06}. Therefore, for the data omission concentrated on specific attributes, it can be identified as an adaptive attack, and an incentive mechanism can be designed to give more severe punishment to this behavior. Through the selection of incentive model parameters, the motivation of SPs to implement adaptive attacks can be reduced.


\begin{figure*}[htbp]
\centerline{\includegraphics[width = 1.0\textwidth]{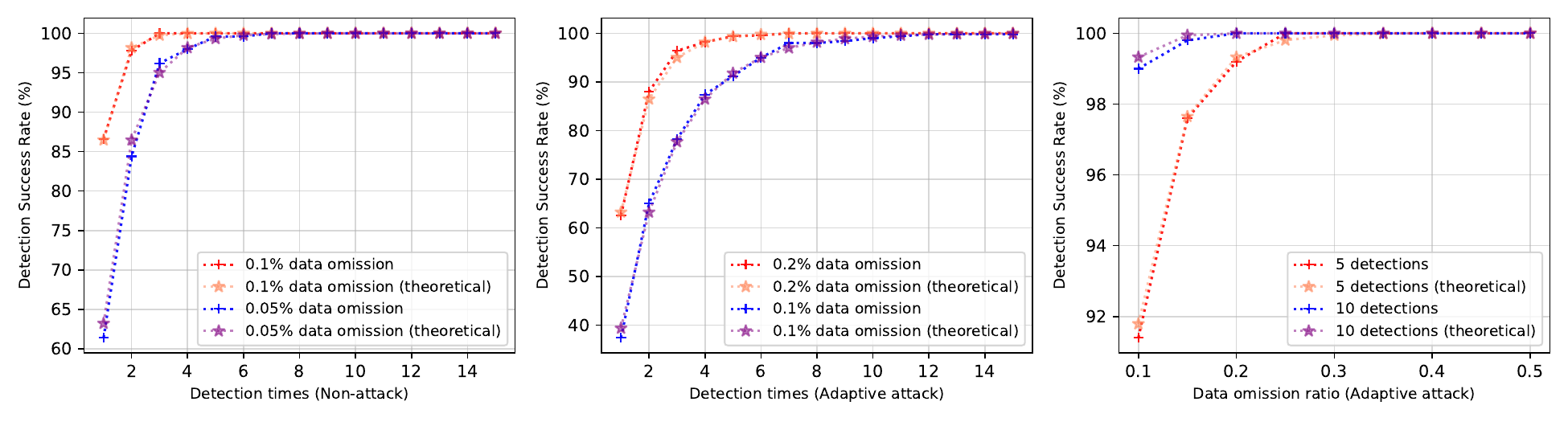}}
\caption{Detection rate of \sv}
\label{exp}
\end{figure*}

\section{Security Analysis} \label{sec:sec}

In this section, we analyze the security of \sv\ and illustrate the advantages of our scheme by analyzing the possible attacks.

According to the threat model mentioned in Section \ref{sec:prob}, the SP may maintain incorrect ADS by omitting or misplacing data objects. Next we provide a formal definition of this threat model and prove the sercurity of \sv.




\textbf{Theorem 2.} In the proposed \sv, when the parameters are properly selected, the probability for SPs to succeed in the following game is negligible:

Run the integrity detection protocol and verifiable query protocol, the SP holds the ADS that supposed to be maintained $\mathcal{A}$, actually maintained ADS $\mathcal{A'}$, \{$Q_d$, $R_d$, $V_{R_d}$\} in the ADS verification protocol.
We say the SP wins the game if $V_{R_d}$ pass the verification and the following condition is satisfied:

\emph{There exists a data object $o_x$ and a query $Q_x$ such that $o_x$ satisfies $Q_x$ when querying $\mathcal{A}$ but $o_x$ does not satisfy $Q_x$ when querying $\mathcal{A'}$.}



\begin{proof}When the SP updates the maintained ADS, the full nodes conduct sampling-based auditing for the ADS. When the SP omit/misplace some data objects in the ADS, he cannot construct valid proof for queries containing the omitted/misplaced data objects. This ensures the correctness of the ADS with high probability. We conduct the proof by contradiction.

\emph{\textbf{Case 1:} The SP maintains incorrect ADS but does not cheat throughout the ADS verification protocol.}

In this case, the full nodes determine whether the SP maintains the correct ADS by verifying the correctness of the returned results. Based on the analysis of attack success probability in Section \ref{sec:sch}, given the security parameter $\lambda$, we can find the proper value of $N$ so that the probability of attacking successfully is no more than $2^{-\lambda}$. Under this condition, we consider the probability of a successful attack to be negligible.


\emph{\textbf{Case 2:} The SP maintains incorrect ADS and cheats by forging the correct result of the detecting token so as to deceive the full nodes.}

In this case, given that there are omitted/misplaced data objects queried by the detecting token, the SP cannot construct proof for the forged result based on $\mathcal{A'}$. If the SP attempts to deceive the full nodes, he must construct an ADS whose root hash is the same as the original one. This suggests the presence of a collision in the cryptographic hash function, which leads to a contradiction.
\end{proof}

\section{Experimental Evaluation} \label{sec:exp}

In this section, we perform experimental evaluation on the proposed \sv\ to prove its effectiveness.

\subsection{Experimental Setting}

We run the experiments on a desktop with Intel(R) Core(TM) i7-12700H 2.3GHz CPU and 16GB memory. We implement the proposed method with Python 3.10, and utilize SHA-256 function for hash calculations. Also, we construct a synthetic dataset with three attributes. The dataset includes a numerical attribute uniformly distributed between 0 and 1,000,000, a discrete attribute and a set attribute containing several keywords (used to simulate keyword queries).

\begin{table}
\centering
\renewcommand{\arraystretch}{1.2}
\caption{Time Cost for ADS Verification (s)}
\vspace{4pt}
\label{tab:timecost}
\begin{tabular}{|c|c|c|c|}
\hline
\multirow{2}{*}{\textbf{Supported query types}} & \multicolumn{3}{c|}{\textbf{Evaluated schemes}} \\
\cline{2-4}
& \multicolumn{1}{c|}{\textbf{vChain+}} & \multicolumn{1}{c|}{\textbf{Baseline}} & \multicolumn{1}{c|}{\textbf{\sv}} \\
\hline
(1) only & 121.9 & 3.41 & 0.269 \\
\hline
(1) and (2) & 243.6 & 10.7 & 0.652 \\
\hline
(1), (2), and (3) & 663.4 & 182.1 & 13.0 \\
\hline
\end{tabular}

\vspace{7pt}
(1) Range-only queries; (2) Object-range joint queries; (3) Keyword queries \& keyword-range joint queries. 
\end{table}

\textbf{Methods to compare. }In existing works\cite{DBLP:conf/icdcs/DaiXYWCH020, DBLP:conf/icde/WangXZXPP22, DBLP:journals/tpds/WuPGYX22} the full nodes directly verify the ADS maintained by the SP by reconstructing them. In order to have a comprehensive understanding of the performance of \sv, we migrate the methods from existing works to our system model and define the following two methods for comparison: 
(1) \textit{Baseline.} The SP constructs MB-tree-based ADS on each combination of attributes, and the full nodes reconstruct the ADS maintained by the SP at certain time intervals to verify its correctness.
(2) \textit{vChain+}\cite{DBLP:conf/icde/WangXZXPP22}. The SP constructs the ADS in \cite{DBLP:conf/icde/WangXZXPP22} for verifiable query, and the full nodes reconstruct the ADS maintained by the SP to verify its correctness.

\textbf{Metrics. }We evaluate the following metrics: (1) \emph{Detection rate:} We measure security of \sv\ with detection rate, which represents the probability of being detected when the SP attempts an attack. (2) \emph{ADS verification cost:} We measure the ADS verification cost with the computational time spent by full nodes to verify the correctness of the ADS. (3) \emph{Query time cost:} We measure the query time cost with the computational time spent by the SP for query and proof construction, and by user for result verification.

\begin{figure}[t]
\centerline{\includegraphics[width = .5\textwidth]{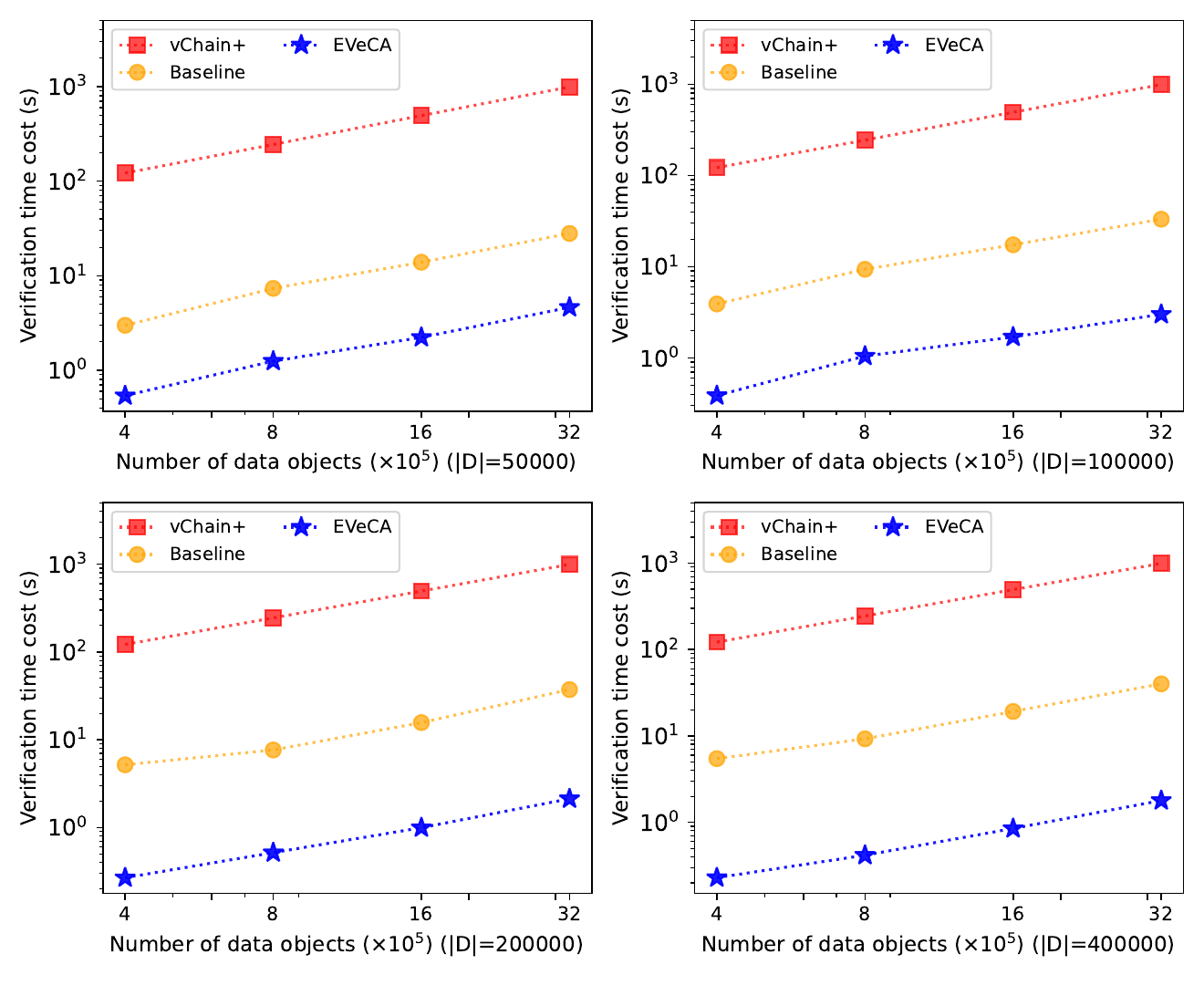}}
\caption{ADS verification time cost for range query}
\label{exp2}
\end{figure}

\begin{figure}[t]
\centerline{\includegraphics[width = .5\textwidth]{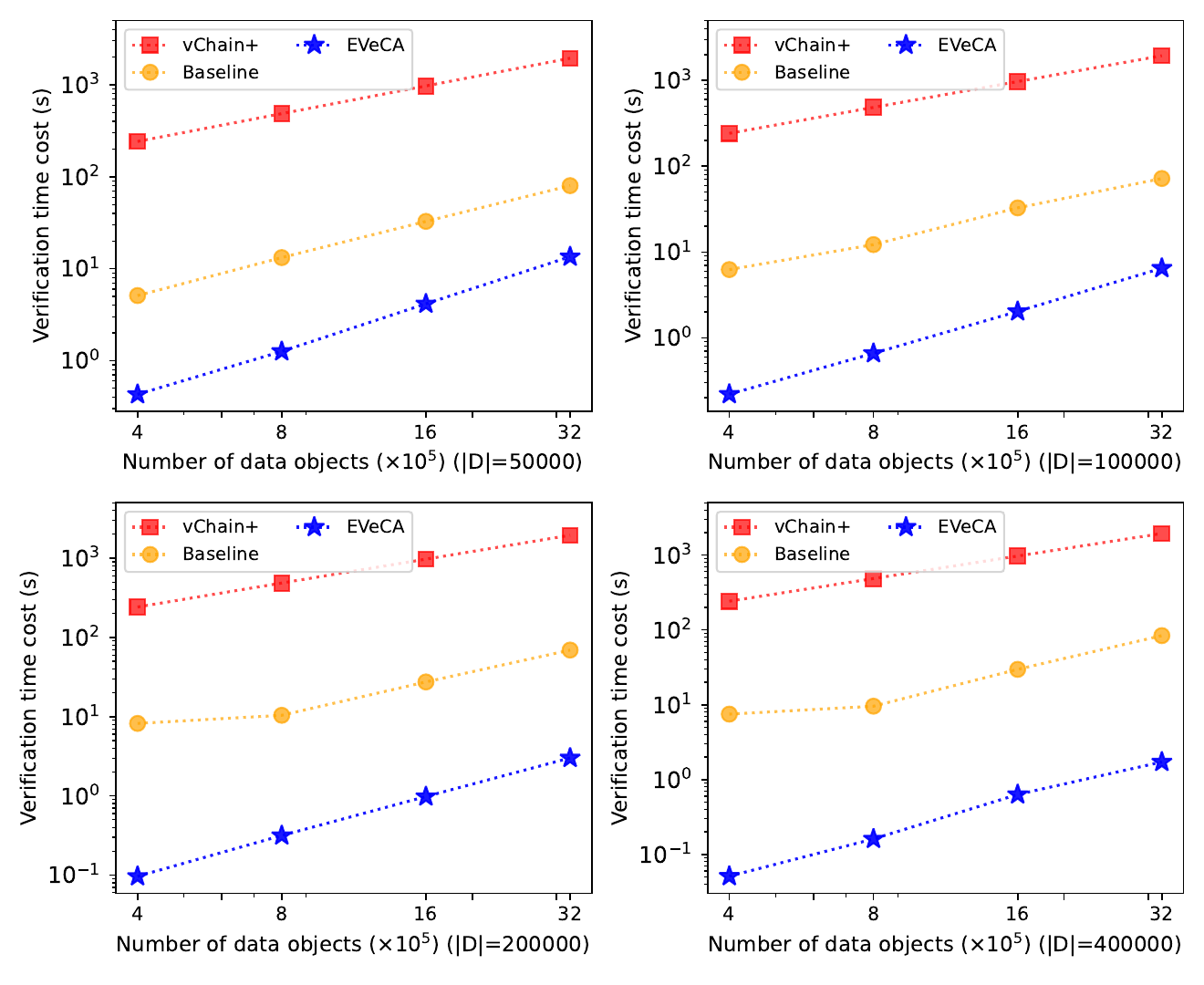}}
\caption{ADS verification time cost for object-range joint query}
\label{exp3}
\end{figure}

\begin{figure}[t]
\centerline{\includegraphics[width = .5\textwidth]{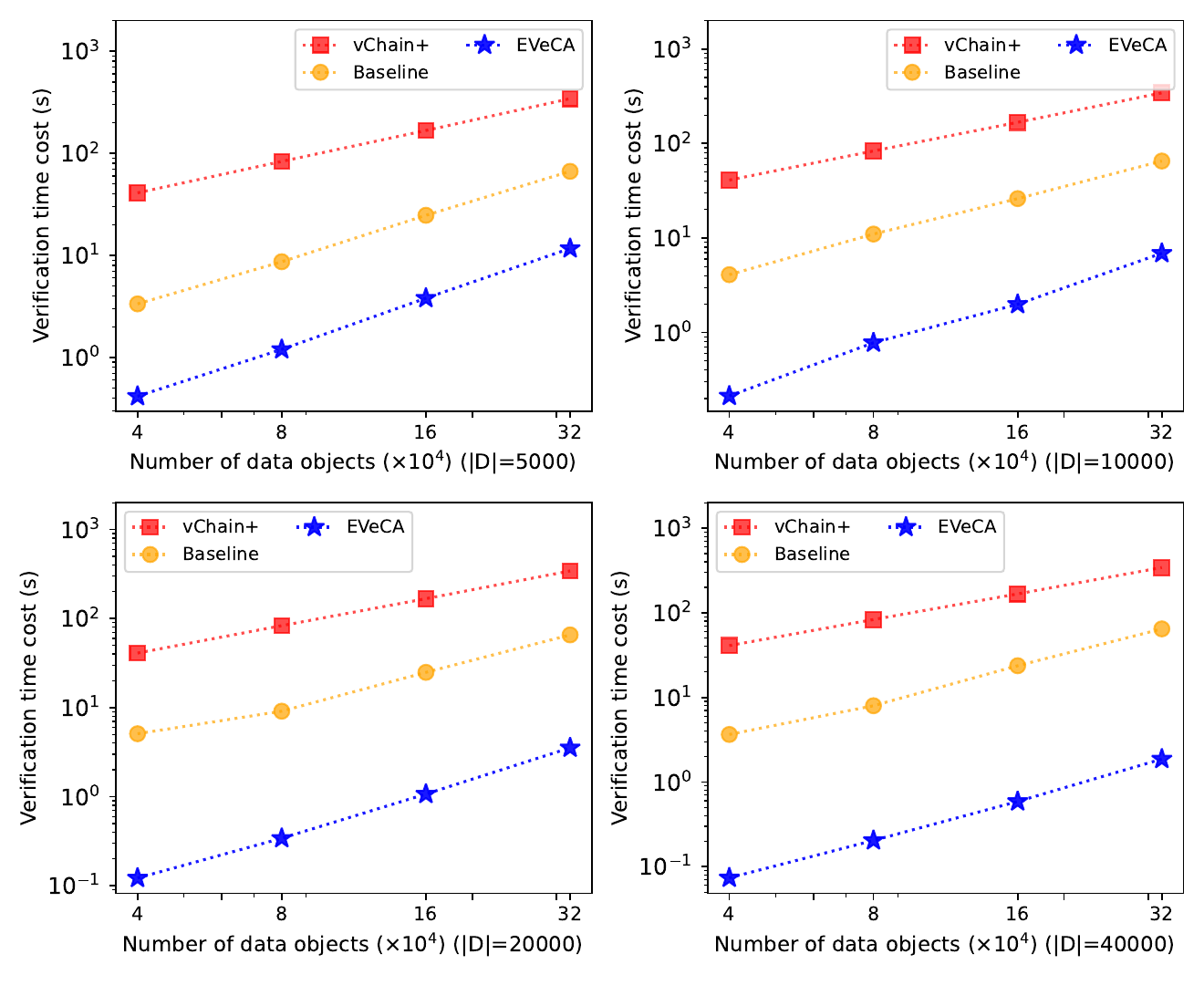}}
\caption{ADS verification time cost for keyword-range joint query}
\label{exp4}
\end{figure}

\subsection{Detection Rate}

We conduct experiments to evaluate the detection rate of \sv. Note that the baseline schemes ensure security and are not within the scope of the current discussion.

\textbf{Non-attack. }We first conduct simulation experiments on the detection success rate of data omission in a single data update. We simulate data omission by randomly deleting a proportion of data objects from the original ADS. The data update size is set to 200,000. We conduct the experimental evaluation with relatively 0.1\%, and 0.05\% data omitted/misplaced by the SP. We evaluate the detection success rate by varying the fraction of data contained in the query expressions. For each set of parameters, we perform 1000 simulations and obtained the results shown in Fig.~\ref{exp}.
The results are align with theoretical derivations.


\textbf{Adaptive attacks. }In the case of adaptive attacks, the challenge node splits the detection expression into an amount of detection ranges and perform the detection. The data update size is set to 200,000. We conduct the experimental evaluation with relatively 0.2\%, and 0.1\% data omitted/misplaced by the SP. When detecting 1\% of the data in a single instance, approximately 10 detections are required to ensure that the detection success rate for missing 0.1\% of the data is not less than 99\%. This aligns with theoretical derivations.

\begin{figure}[t]
\centerline{\includegraphics[width = .5\textwidth]{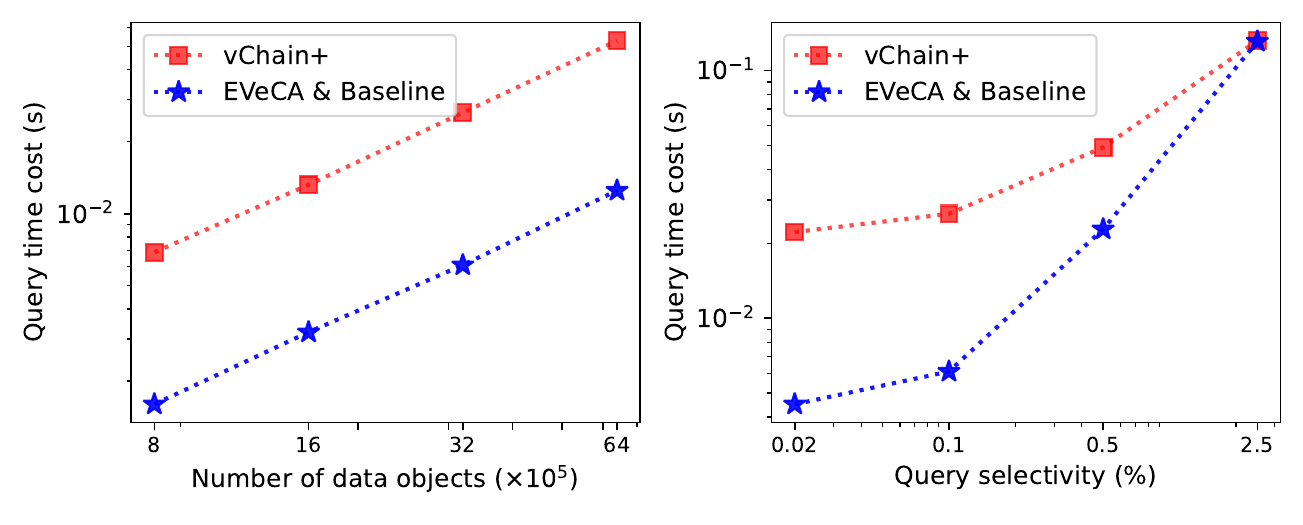}}
\caption{Query time cost on range query}
\label{exp5}
\end{figure}

\begin{figure}[t]
\centerline{\includegraphics[width = .5\textwidth]{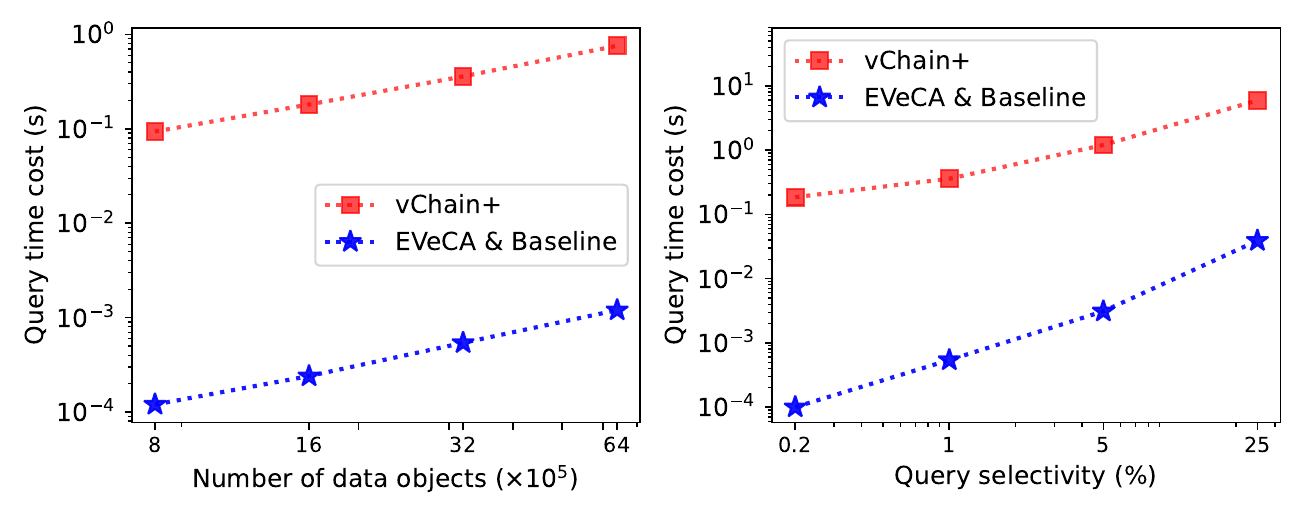}}
\caption{Query time cost on object-range joint query}
\label{exp6}
\end{figure}

\begin{figure}[t]
\centerline{\includegraphics[width = .5\textwidth]{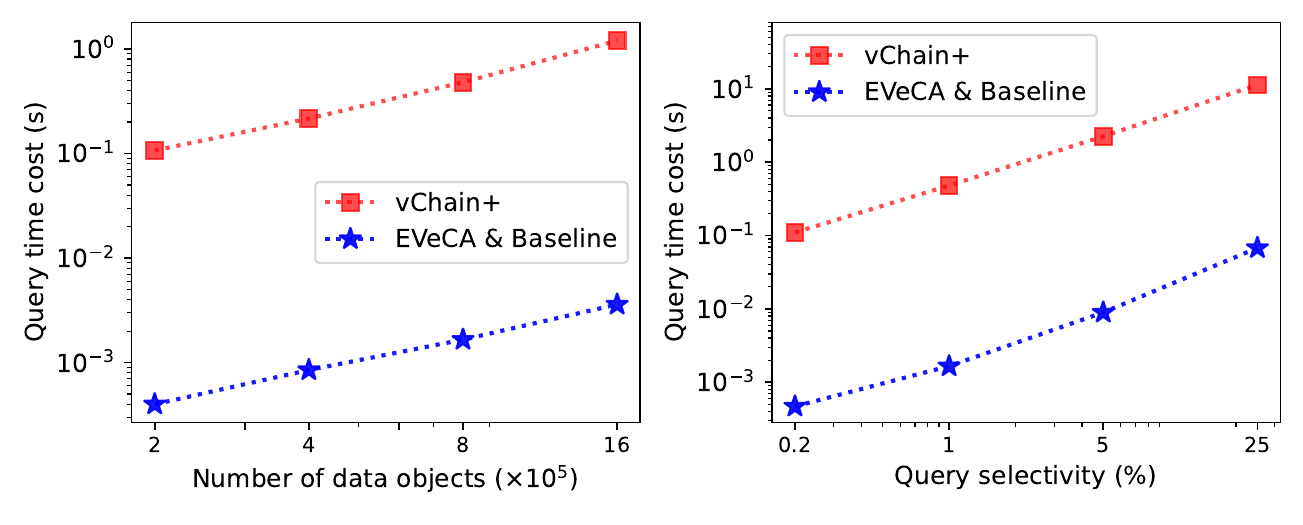}}
\caption{Query time cost on keyword-range joint query}
\label{exp7}
\end{figure}

\subsection{ADS Verification Cost}


We conduct a series of comparative experiments with the \sv\ and the baseline schemes. We all adopt our constructed synthetic datasets to evaluate the respective efficiency of the two methods. We test the computational time cost of the full nodes to verify the query result. The supported query types include: (1) Range-only queries; (2) Object-range joint queries; (3) Keyword queries \& keyword-range joint queries. 

We conduct experiments on ADS verification overhead under different combinations of query types shown in Table~\ref{tab:timecost}. The number of data objects is set to 400,000. It can be observed that as the number of supported multi-attribute query types increases, the rate at which vChain+ incurs time cost is lower compared to Baseline and \sv. This is expected as the multi-dimensional queries in vChain+ utilize an intersection-based approach across dimensions, which eliminates the overhead of constructing composite indexes. In contrast to the Baseline, \sv\ achieves efficiency gains of 10-20 times.



\textbf{Range-only queries. }We set the number of data objects as $[4, 8, 16, 32]\times 10^5$ in turn. The maximum tree size is set to 200,000. From the results, we can see that our scheme is about 20 times more efficient than the baseline scheme. We set the update size as $[5, 10, 20, 40]\times 10^4$ and conduct experiments separately. As shown in Fig.~\ref{exp2}, the verification cost decreases gradually as $|D|$ increases. This aligns with the pattern presented in the earlier formula derivations.


\textbf{Object-range joint queries. }The ADS validation overhead for object-range queries is depicted in Fig.~\ref{exp3}. We take the numerical attribute and the discrete attribute of our constructed data. We build composite index of the test data by organizing them in a forest with a set of Merkle B+ trees. We set the number of data objects as $[4, 8, 16, 32]\times 10^5$ and update size as separately $[5, 10, 20, 40]\times 10^4$. The verification cost decreases as $|D|$ increases.

\textbf{Keyword-range joint queries. }For keyword-range queries, we adopt the scheme of Merkle inverted index to build an MB tree on each keyword as ADS. The number of keywords is set to 400 and each data object randomly contains 2-20 keywords. We set the number of data objects as $[4, 8, 16, 32]\times 10^4$ and the update size as separately $[5, 10, 20, 40]\times 10^3$. As shown in Fig.~\ref{exp4}, the verification cost also decreases as $|D|$ increases.




\subsection{Query Performance}

The comparison in \sv\ and the baseline schemes on query performance are shown in Fig.~\ref{exp5} to Fig.~\ref{exp7}. We conduct evaluations by varying the number of data objects and query selectivity. We consider range-only queries, multi-dimensional range joint queries and keyword-range joint queries. Note that since Baseline shares the same method of constructing ADS as \sv\, thus resulting in the same query cost for both. For range-only queries, \sv\ performs better when the selectivity is below 2.5\%. This is mainly because \sv\ has a larger ADS size, which reduces the time overhead. However, since the time overhead of range queries based on the MB-tree is linearly related to the size of the query results, our approach no longer exhibits clear advantages as the selectivity increases.
For both types of 2-dimensional queries, \sv\ has clear advantages over vChain+. This is because \sv\ utilizes composite ADS, while vChain+ is limited by one-size-fits-all ADS, and involves querying each attribute separately and then taking the intersection.

\section{Related Work}

In this section, we take a review of important works in recent years. We begin with verifiable queries of databases, then introduce verifiable query on blockchain. Lastly, we summarize the limitation of existing works.

\subsection{Verifiable Query for Databases}

Many works have been proposed to solve the trust problem of outsourced databases with verifiable query, where the data owner outsources the database to a cloud server, and the cloud server provides verifiable query services to users. Merkle B-tree (MB-tree)\cite{DBLP:conf/sigmod/LiHKR06} combines Merkle tree with B+ tree, which enables efficient query on one-dimensional data. Most of the works use the Merkle-tree-based method to deal with verifiable query problem\cite{DBLP:conf/dbsec/Nuckolls05, DBLP:conf/icde/PangT04, DBLP:journals/tissec/LiHKR10}, and there also exists works that use probabilistic method\cite{DBLP:conf/vldb/XieWYM07, DBLP:journals/jcs/VimercatiFJPS16}, where the client insert fake data tuples into the database, and check whether the SP return the correct answer on fake tuples. For dishonest-data owner model, DE-Sword\cite{DBLP:journals/tdsc/QiZLCMYQ22} introduces a double-edged reputation incentive mechanism enable query verifiability. There are also works that use hardware-based schemes to enable verifiable query\cite{DBLP:conf/sigmod/ZhouCPWM021, DBLP:journals/pvldb/BajajS13}.

There are also works about verifiable queries on database using blockchain. Zhang\etc\cite{DBLP:conf/icde/Zhang0XTC19} propose $GEM^2$-Tree, an ADS that can be efficiently maintained while being effective in supporting authenticated range queries. Similar to $GEM^2$-Tree\cite{DBLP:conf/icde/Zhang0XTC19}, the authors in\cite{DBLP:journals/tifs/WangZDCJ23} design a kind of ADS to support skyline query, and propose the scheme of data update and verifiable query. Peng\etc\cite{DBLP:conf/sigmod/PengDL0S20} propose a blockchain-based collaborative database which support efficient and verifiable query. Then, Zhang\etc\cite{DBLP:conf/icde/ZhangXWXC21} propose a Chameleon inverted index that utilizes vector commitment to reduce ADS maintenance cost.

The above works on outsourced database verifiable queries are difficult to migrate to on-chain data querying scenarios for the following reasons. First, in contrast to the traditional scenario where the ADS is managed by the data owner, in on-chain data scenarios where there is no trusted data owner, the responsibility of maintaining the ADS falls upon the full nodes. Due to the high-frequency updates of on-chain data, the maintenance of ADS will impose a heavy burden on full nodes. Second, in outsourced databases, it is always possible to generate and add new ADS as required to facilitate additional queries. However, this becomes challenging due to the immutability of the blockchain.

\subsection{Blockchain Data Querying}

Due to the inefficiency of blockchain data query, many works focus on data management on the blockchain. One approach is to extract the on-chain data to off-chain databases. In EtherQL\cite{DBLP:conf/dasfaa/LiZYLZ17}, the on-chain data is extracted to databases to support efficient queries. Due to the need to ensure the reliability of the query, VQL\cite{DBLP:journals/tpds/WuPGYX22} propose middleware layer to build micro databases, and utilize Merkle Patricia tree to manage fingerprints and put forth a simplified query result verification algorithm for users. However, this scheme fails to ensure the completeness of the query result.

Another common method is to construct on-chain ADS to enable verifiable on-chai data queries. In this type of approach, the users maintain the digest of the on-chain index as reliable prior knowledge. Xu\etc\cite{DBLP:conf/sigmod/0004ZX19} propose vChain, a verifiable query scheme for on-chain data which supports boolean range queries. Zhu\etc\cite{DBLP:conf/icde/ZhuZJZY19} build an on-and-off-chain index structure to support multiple types of queries and implement rich authenticated queries on the blockchain for light client. LVQ\cite{DBLP:conf/icdcs/DaiXYWCH020} introduces an ADS of on-chain data based on Bloom filter integrated Merkle tree, which realize lightweight verifiable queries. Wang\etc\cite{DBLP:conf/icde/WangXZXPP22} build a fixed-length sliding window index on the basis of vChain\cite{DBLP:conf/sigmod/0004ZX19} to reduce the query time overhead at the cost of introducing a large consensus overhead. Yin\etc\cite{DBLP:journals/tsusc/YinZZW23} propose a kind of ADS based on bloom filter and Merkle R-tree that supports efficient data query and result authentication.

Despite the aforementioned works on verifiable queries in blockchain, they have limitation in common. They rely on the on-chain ADS to support verifiable queries, and are difficult to balance between query efficiency and index maintenance efficiency for data with multiple attributes.

\section{Conclusion}

In this paper, we propose EVeCA, a novel framework that addresses the challenge of efficient and verifiable on-chain data queries. EVeCA delegates the tasks of ADS maintenance and query service provision to a limited number of service providers (SPs), thereby reducing the computational burden on full nodes. The correctness of the ADS maintained by SPs is ensured through a challenge-based authentication scheme. In this scheme, a challenge node generates a detecting token with known results and sends it to the SP as a challenge. Thus, the full nodes can determine the correctness of the maintained ADS by verifying the returned result of the detecting token. Based on the ADS verified by full nodes, users can further perform verifiable query processes with the SP. Our framework significantly reduces the need for full nodes to maintain on-chain ADS, enabling them to participate in ADS maintenance at minimal cost. As for future work, we plan to incorporate privacy protection requirements and migrate our solution to the scenario of encrypted data.

	\bibliographystyle{IEEEtran}
        \renewcommand{\baselinestretch}{1.25}
        \bibliography{refs}

\end{document}